\documentclass[reprint,superscriptaddress,
amsmath,amssymb,aps,
]{revtex4-2}
\usepackage{graphicx} 
\usepackage{lineno}
\usepackage[colorlinks=true,
	linkcolor=blue,
	filecolor=blue,      
	urlcolor=blue,
	citecolor=blue,]{hyperref}
\begin{document}
\title{Quantum light and radiation in Rindler spacetime: from uncertainty relations to the cosmological implications}

\author{Fujin Wang} 
\affiliation{School of Physics, Zhejiang University, Hangzhou 310058, China}

\author{Syed Masood}
\affiliation{Department of Physics, School of Science and Research Center for Industries of the Future, Westlake University, Hangzhou 310030, P. R. China}
\affiliation{Institute of Natural Sciences, Westlake Institute for Advanced Study, Hangzhou 310024, P. R. China}

\author{L.G. Wang}
\email{lgwang@zju.edu.cn}
\affiliation{School of Physics, Zhejiang University, Hangzhou 310058, China}
\thanks{Corresponding author}

\date{\today}
\begin{abstract}   
{Based on an analogy between diffraction integral formalism of classical field propagation and Feynman path integral approach to quantum field theory, we develop a quantum model for light and radiation in Rindler spacetime. The framework helps to reveal acceleration-induced contributions to the traditional Heisenberg position-momentum uncertainty relation. A modified Planck energy density distribution of radiation is established and reveals equivalence between temperature and Rindler acceleration as advocated by standard Unruh and anti-Unruh effects. Later, by defining an equivalent acceleration, we investigate some cosmological implications of the model with regards to redshift and expansion of the Universe. In this context, we contend that the accelerated expansion of the Universe, in addition to possessing some well-defined limits corresponding to early and local Universe epochs, may also hint towards dynamical nature of dark energy. The findings provide glimpse into future table-top experiments aimed at emulating gravitational and other cosmological phenomena in terrestrial lab setups.}
\end{abstract}


\maketitle

\section{Introduction}
Following the advent of general relativity by Einstein and other alternative approaches to gravity \cite{hollands2015}, the study of field propagation on curved geometries has been a subject of great fascination \cite{capozziello2011}. It spans a wider class of approaches with the field being treated either on classical or quantum footings. Whether the background for the field propagation is the three-dimensional curved surface or four-dimensional spacetime of gravity, the results have been pivotal in advancing our understanding of behavior of the fields subjected to arbitrary external circumstances. This, in part, reflects the ability of fields to offer a potential window into the workings of the Universe. Furthermore, the idea of understanding the electromagnetic field (light) propagation in curved spaces is in line with the spirit of emulating general relativistic phenomena in controlled lab settings \cite{bekenstein2015}. 

Classical approaches have been by far insightful in many aspects. For example, Maxwell's equations in curved space reveal the coupling mechanism between electromagnetic field and the Riemann curvature tensor, and the systems can be solved using nonlinear Schr\"{o}dinger equation \cite{batz2008}. Using analog gravity systems \cite{barcelo2011}, it is possible to model the nonlinear nature of gravity using optical wavepackets, with the resulting system dynamics being analogous to Newton-Schr\"{o}dinger approach \cite{bekenstein2015}. Furthermore, optical studies of two-dimensional curved surfaces in this regard has highlighted many interesting facets of role of geometry in shaping the propagation dynamics of light \cite{xu2021,zhang2024,ding2021a}. 
   
The quantum aspects of field propagation in curved spacetime typically involve massless scalar field as the widely used instructive model owing to its mathematical and physical simplicity \cite{capozziello2011}. A myriad of phenomena involving Hawking radiation from black holes \cite{hawking1975}, Unruh effect in Rindler spacetime \cite{rindler1966,unruh1976,unruh1984,gerlach1976}, Parker particle creation in expanding spacetime \cite{parker1968,parker1969,parker1971}, moving mirror radiation \cite{Fulling1976,akal2021}, or dynamical Casimir effect \cite{lock2017}, is a witness to that.  Electromagnetic field has also been a source of physical insights beyond scalar field approximation, particularly in probing the geometric structure of black holes \cite{bambi2017}. Although, most of these studies rigorously make use of standard quantum field theoretic approach in curved spacetime \cite{capozziello2011}, however, the possibility of using simpler and handy frameworks which have further revealed the rich field theoretic aspects of curved spacetime, have recently been pursued. For example, by maintaining field in a vacuum state within an accelerated cavity in black holes, freely-falling atoms through this cavity give rise to a surprising Hawking-like radiation emission \cite{scully2018}, which otherwise does not arise under same conditions in the usual Hawking picture \cite{hawking1975}. One key point related to the cosmological aspects of light propagation under realistic scenarios is the need for taking into consideration the enormous geometric extent of the astrophysical objects. The reason for this stems from the fact that curved spacetime effects are generally perceptible over longer distances, which is one of the fundamental implications of general relativistic idea of gravity being curvature of the spacetime geometry \cite{Carroll2019}.\\
\indent This, along with several other factors, renders direct observation of quantum radiation associated with real curved spacetime of the Universe extremely challenging, thereby strengthening the case for the analogue gravity paradigm \cite{barcelo2011}. In particular, it has been possible to observe light-deflection in Schwarzschild-like geometry by invoking nanostructured thin films \cite{bekenstein2017}, while as reconfigurable meta-surfaces have helped us to realize wormhole-like topological channels \cite{zhu2018}. Furthermore, gradient-index media are also widely used to mimic gravitational lensing effects \cite{sheng2013,xiao2021}. Another interesting avenue involves use of Bose-Einstein condensates in acoustic black holes to study Hawking radiation \cite{faccio2013,munozdenova2019}. For further delving into this fascinating field, we refer the interested reader to Refs. \cite{hartle1983,ratzel2018,adesso2007,kish2016,bornman2019,kish2022,ding2022}. 

Inspired by the advances made in these interdisciplinary research fields, we are motivated to develop a model for light and radiation in accelerated (Rindler) frame with a blend of two major approaches to light dynamics: diffraction formula for light radiation and Feynman path integral approach \cite{ding2021b,ding2022,feynman2010}. The existing duality between these two approaches is justified from the universal nature of Green function kernel involved in the mathematical representation of the two theories for description of field evolution. After conceptualizing the model, we investigate acceleration-induced effect on the wavepacket dynamics of light from the perspective of an observer in Rindler spacetime while assuming wavepacket to be that of free quantum particle. We demonstrate how acceleration effects are manifested in wave dispersion and uncertainty relations. It is further argued that this model offers an opportunity to explain cosmological redshift mechanism in Lambda-cold-dark-matter ($\Lambda$CDM) model of cosmology \cite{abdalla2022} by assuming an equivalent acceleration of the Universe in analogy with the expanding geometry of Friedmann-Lemaître-Robertson-Walker metric (FLRW) metric \cite{bergström2006,young2017}. This way we aim to bridge quantum field effects in Rindler spacetime to the expanding Universe, with the hope to make any relevant case for future constraining of various cosmological parameters of the Universe.

\section{Building the theory}\label{sec2}
\subsection{Collins diffraction formula and quantum path integral approach: an analogy}
Here, we draw an analogy between classical wave picture of a field and quantum field theoretic perspective. We begin from the metric of the two-dimensional Rindler spacetime expressed as $\mathrm{d}s^2 = e^{2a z/c^2} \cdot \left(-c^2 \mathrm{d}t^2 + \mathrm{d}z^2\right) + \mathrm{d}x^2$, which describes a non-inertial frame with constant acceleration $a$ along $z$ axis. Under fixed temporal coordinate $t$, the behavior of light in Rindler spacetime can be characterized by diffraction integrals. Notably, the Collins diffraction formula governing such evolution of light has been analytically derived under paraxial approximation \cite{ding2021b}, expressed as 
\begin{widetext} 
\begin{align} 
    E(x_2,z_2) 
    =&\sqrt{ -\frac{ika/c^2}{\pi (e^{2a(z_2-z_1)/c^2} - 1)} } \,  \int\limits_{-\infty}^{\infty} E(x_1,z_1) \exp\left\{ \frac{ \frac{ika}{c^2}\left( x_1^2 - 2x_1 x_2 + e^{2a(z_2-z_1)/c^2}\, x_2^2  \right)}{  e^{2a(z_2-z_1)/c^2} - 1 } \right\} \, \mathrm{d}x_1 \nonumber\\
    \approx& \sqrt{ -\frac{ika/c^2}{\pi (e^{2a(z_2-z_1)/c^2} - 1)} } \,  \int\limits_{-\infty}^{\infty} E(x_1,z_1) \exp\left\{ \frac{ \frac{ika}{c^2} \left( x_1^2 - 2x_1 x_2 + x_2^2 \right) }{ e^{2a(z_2-z_1)/c^2} - 1 } \right\} \, \mathrm{d}x_1,
    \label{eq1}
\end{align}
\end{widetext}
where $c$ denotes the vacuum speed of light, $k$ is the wave number of light. Here, $E(x_1, z_1)$ is the input optical field and $E(x_2, z_2)$ is the output optical field.

The Feynman path integral formulation is a fundamental description of quantum mechanical evolution of fields \cite{feynman2010}. The probability amplitude of the state of a particle is sum over all possible paths that the particle would take between two points. This approach in some way bridges classical and quantum mechanics, finding extensive applications in quantum optics, quantum field theory, and cosmology \cite{feynman1948,hartle1983}. The formalism involves time evolution of a wavepacket being described by so-called propagators. In general, the mathematical description of the phenomenon is given by 
\begin{equation}
    \Psi(x_2, t_2) = \int \Psi(x_1, t_1) Q(x_2, t_2; x_1, t_1)\mathrm{d}x_1,
    \tag{2a}
    \label{eq2a}
\end{equation}
where $Q$ is the propagator, defined as  $Q(x_2, z_2; x_1, z_1)\sim\exp\{iS_{cl}(x_2, z_2; x_1, z_1)/\hbar\}$ with $S_{cl}$ as classical action relating two points in spacetime. Note that this is somehow the advanced form of unitary time evolution operator in Schrodinger wave formalism. Now, evoking Collins approach, the evolution amplitude of light between two points $(x_1, z_1)$ and $(x_2, z_2)$ is generally governed by the transformation
\begin{equation}
E(x_{2},z_{2}) = \sqrt{ -\frac{ik}{2\pi B} } \int E(x_{1},z_{1}) \, e^{ikL} \, dx_{1}.
\tag{2b}\setcounter{equation}{2}
\label{eq2b}
\end{equation}
A closer look at Eqs.~\eqref{eq2a} and \eqref{eq2b} reveals a striking resemblance between the two approaches, particularly with the involvement of time evolution operator in the former and eikonal phase factor in the latter. 

The analogy thus established through Eqs.~\eqref{eq2a} and \eqref{eq2b} demonstrates that the light propagation distances $z_{1,2}$ in optical fields using Collins approach are formally equivalent to the evolution times $t_{1,2}$ of the wave function in path integral formalism, with their relationship fundamentally connected by the "velocity." Additionally, the light distribution $x$ corresponds to the wave function distribution, and the eikonal function are analogous to the phase of the propagator. These equivalences suffice to lead us to propound that the propagator in Rindler spacetime is expressed as
\begin{align}
Q&= \sqrt{ -\frac{ika/c^2}{\pi (e^{2a(z_2-z_1)/c^2} - 1)} } \nonumber\\
&\quad \times \exp\left\{ \frac{\frac{ika}{c^2}\left( x_1^2 - 2x_1x_2 + x_2^2 \right)}{e^{2a(z_2-z_1)/c^2} - 1}  \right\},
\label{eq3}
\end{align}
where $Q$ represents the propagator $Q(x_2,z_2;x_1,z_1)$, demonstrates the probability amplitude for the optical field propagating from $(x_1, z_1)$ to $(x_2, z_2)$ in Rindler spacetime. Furthermore, within the Feynman path integral framework, the propagator can also be expressed as a linear superposition of eigenstates \cite{feynman2010}:
\begin{align}
Q=& \sum_{n=1}^\infty \psi_n^*(x_1) \psi_n(x_2) e^{-\frac{iE_n(t_2-t_1)}{\hbar}}\nonumber \\
=& \int \psi_p^*(x_1) \psi_p(x_2) e^{-\frac{iE_p(t_2-t_1)}{\hbar}} \mathrm{d}p \nonumber \\
=& \int \frac{1 }{2\pi\hbar}\exp\{ \frac{ip(x_2-x_1)}{\hbar}\}\nonumber\\
&\times \exp\{- \frac{ ip^2(e^{2a(z_2-z_1)/c^2}-1)}{(4ak\hbar^2)/c^2} \}\mathrm{d}p.
\label{eq4}
\end{align}
When the quantum number $n$ is too large, the spectrum becomes continuous, as is the case with a free particle, the summation over $n$ reduces to an integration over momentum $p$, where $E_n$ denotes the energy of the eigenstate $\psi_n(x)$. It illustrates that the spatial distribution of the optical field is mathematically equivalent to the distribution of a free-particle wave function, and its propagation also corresponds to the evolution of a free-particle wave function. Furthermore, if we assume that the effective propagation distance of the optical field is given by $z_{eff} = \frac{ \exp\left\{2a(z_2 - z_1)/c^2\right\} - 1 }{2a/c^2}$, we can deduce the equation for effective velocity as 
\begin{equation}
\frac{p^2 z_{\text{eff}}}{2k\hbar^2} = \frac{k v'(t_2 - t_1)}{2},
\label{eq5}
\end{equation}
where momentum $p=k\hbar$, $v'$ denotes the group velocity of optical field in Rindler spacetime, and $(t_2-t_1)$ corresponds to the propagating time in lab frame. With this in hand, we can establish relation between the propagating distance and group velocity in flat spacetime. By noting that $z_2-z_1 = v(t_2-t_1)$, thus we have the following expression,
\begin{equation}
v' = \frac{e^{2a(z_2 - z_1)/c^2} - 1}{2a(z_2 - z_1)/c^2} v,
\label{eq6}
\end{equation}
which reveals a fundamental distinction between the behavior of group velocity of optical field in Rindler spacetime and in flat spacetime. Remarkably, the group velocity becomes analogous to the dynamical velocity of a quantized free-particle model. By setting the initial position to zero $(z_1=0)$, Eq.~\eqref{eq6} yields as
\begin{equation}
v' = \frac{e^{2a z/c^2} - 1}{2a z/c^2} v = \frac{e^{2\Lambda} - 1}{2\Lambda} v,
\label{eq7}
\end{equation}
where the parameter $\Lambda = az/c^2$ quantifies the strength of acceleration effects associated with Rindler spacetime. This relationship demonstrates that light propagation in Rindler spacetime is mathematically equivalent to whole wavepacket undergoing uniform acceleration in flat spacetime, which is the essence of Rindler transformation. Consequently, this theoretical equivalence provides a method for experiments to simulate light propagation in Rindler spacetime by modulating the group velocity \cite{hu2019,hall2022}.

It may be noted that the classical action $S_{cl}$ involved in the propagator $Q(x_2, z_2; x_1, z_1)$ of path integral approach is evaluated along the extremal path from an initial point $(x_1, z_1)$ to a final point $(x_2, z_2)$. This corresponds to the least-action in classical mechanics and the optical path length along geodesics in geometrical optics. We use Eq.~\eqref{eq3} to write the classical action as 
\begin{equation}
S_{cl} = \frac{\hbar k}{2v'(t_2 - t_1)} (x_1 - x_2)^2.
\label{eq8}
\end{equation}
The angular frequency is then determined as \cite{feynman2010}
\begin{equation}
\omega' = -\frac{\partial S_{cl}}{\hbar \partial t} = \frac{k v_x^2}{2v'} = \frac{k v'}{2}.
\label{eq9}
\end{equation}
This formulation is based on analogy of the equivalence between the group velocity of the optical field and the velocity of a quantum particle, i.e. $ v' = v_x$. Under this assumption, the angular frequency $\omega'$ of optical field also serves as the angular frequency of a free-particle in Rindler spacetime. Consequently, the corresponding energy for the free-particle is defined as $ E' = \hbar \omega'$. Furthermore, within this customized model, the phase velocity and group velocity become identical, i.e. $ \mathrm{d}\omega'/\mathrm{d}k = \omega'/k$.

\subsection{Wave function evolution and uncertainty relations}
\begin{figure}[htbp]
   \centering
   \includegraphics[width=1\linewidth]{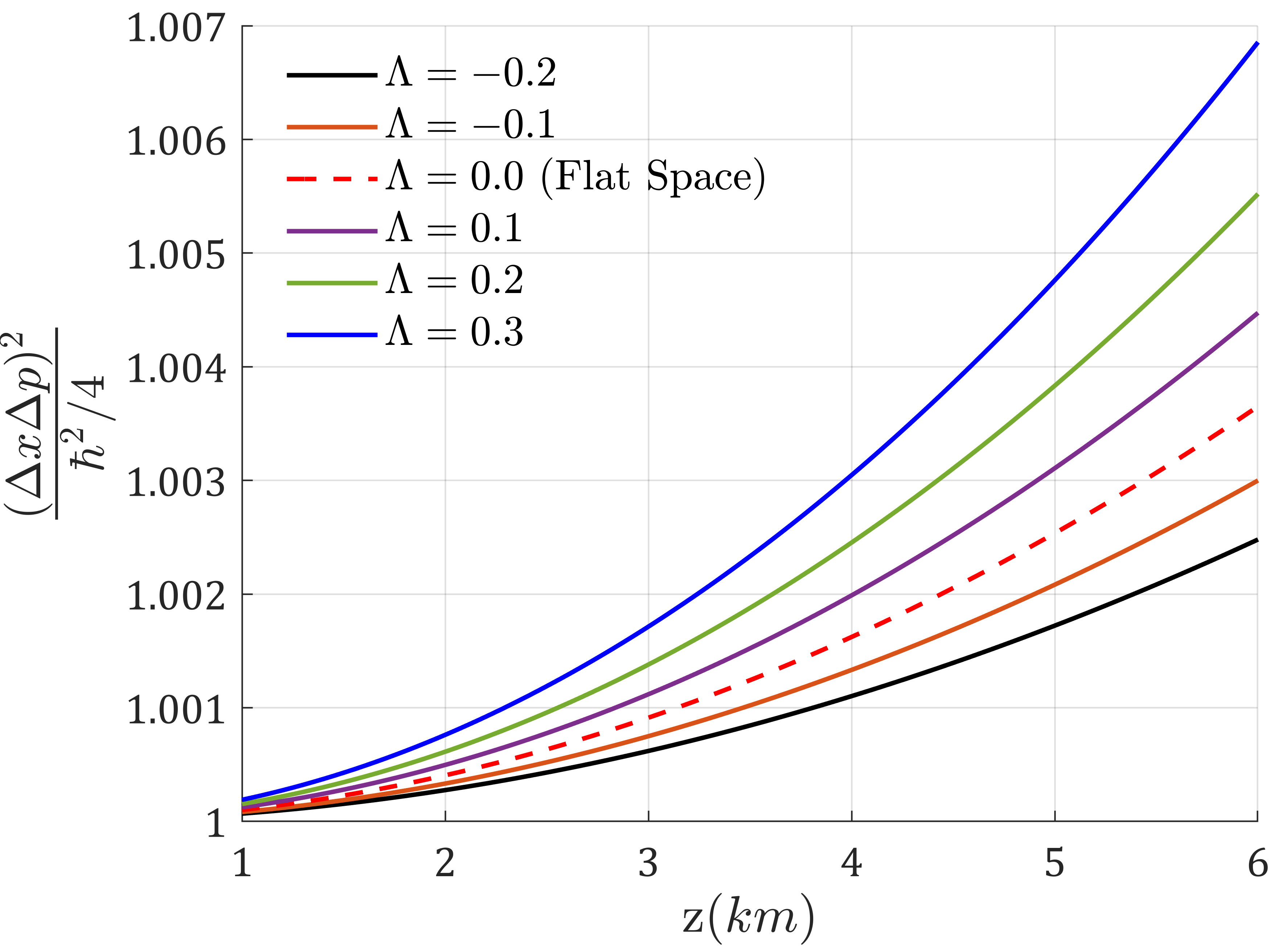}
      \caption{Changes of the momentum-position uncertainty relations. Parameters include $\lambda = 632.8~\mathrm{nm}$, $\sigma_0 = 0.1\mathrm{m}$.}
    \label{Fig1}
   \end{figure}

Following Eq.~\eqref{eq4}, the eigenstates of wave function under coordinate representation are derived as:
\begin{equation}
\psi_p(x,z) = \frac{e^{ipx/\hbar}}{\sqrt{2\pi\hbar}} \exp\left\{ -i \frac{p^2 z}{2k\hbar^2} \cdot \frac{e^{2\Lambda} - 1}{2\Lambda} \right\}.
\label{eq10}
\end{equation}
Though the eigenstates formally resemble plane-wave functions, their evolution in the accelerated (Rindler) frame differs due to a corrective phase factor originating from the parameter $\Lambda$. The complete quantum state is constructed as a linear superposition of these eigenstates:
\begin{equation}
\Psi(x,z) = \frac{1}{\sqrt{2\pi\hbar}} \int \phi(p) \, e^{i\left(\frac{px}{\hbar} - \frac{p^2 z}{2k\hbar^2} \cdot \frac{e^{2\Lambda}-1}{2\Lambda}\right)} \mathrm{d}p.
\label{eq11}
\end{equation}
The form of $\phi(p)$ is uniquely determined by the initial state $\Psi(x,0)$ through the Fourier transform. Let's assume the initial state to be a normalized Gaussian wave packet,
\begin{equation}
\Psi(x,0) = \frac{1}{(\pi \sigma_0^2)^{1/4}} e^{-\frac{(x-x_0)^2}{2\sigma_0^2}},
\label{eq12}
\end{equation}
where $\sigma_0$ represents the initial spatial width of the wave packet, and $x_0$ corresponds to its initial central position. Then $\phi(p)$ can be derived as
\begin{align}
\phi(p) &= \frac{1}{\sqrt{2\pi\hbar}} \int \Psi(x,0) e^{-i px/\hbar} \mathrm{d}x \nonumber\\
&= \frac{\sigma_0}{\sqrt{\hbar} (\pi\sigma_0^2)^{1/4}} \exp\left\{ -\frac{p\left(p + 2i\hbar x_0/\sigma_0^2\right)}{2\hbar^2/\sigma_0^2} \right\}.
\label{eq13}
\end{align}
By substituting $\phi(p)$ into Eq.~\eqref{eq11}, the complete free-particle wave function in Rindler spacetime can be expressed as
\begin{align}
\Psi(x,z) =& \frac{1}{(\pi \sigma_0^2)^{1/4} \sqrt{1 + i \frac{z}{k \sigma_0^2} \cdot \frac{e^{2\Lambda} - 1}{2\Lambda}}}\nonumber \\
&\times \exp\left\{ -\frac{(x-x_0)^2}{2\sigma_0^2 \left(1 + i \frac{z}{k \sigma_0^2} \cdot \frac{e^{2\Lambda} - 1}{2\Lambda}\right)} \right\}.
\label{eq14}
\end{align}
Eq.~\eqref{eq14} satisfies the normalization condition and degenerates into the free-particle wave function in flat space in the vanishing acceleration limit ($a\rightarrow0$), i.e.,
\begin{align}
\Psi(x,z) =& \frac{1}{(\pi \sigma_0^2)^{1/4} \sqrt{1 + i \frac{z}{k \sigma_0^2}}} \nonumber\\
&\times\exp\left\{ -\frac{(x-x_0)^2}{2\sigma_0^2 \left(1 + i \frac{z}{k \sigma_0^2}\right)} \right\}.
\label{eq15}
\end{align}
Now, based on Eq.~\eqref{eq14}, the expectation value of the particle's position and its position uncertainty can be derived as
\begin{align}
\langle x \rangle &= \langle x_0 \rangle,\tag{16a} \\
\langle x^2 \rangle &= \frac{\sigma_0^2}{2} \left[1 + \left( \frac{z}{k\sigma_0^2} \cdot \frac{e^{2\Lambda} - 1}{2\Lambda} \right)^2 \right] + x_0^2,\tag{16b} \\
(\Delta x)^2 &= \frac{\sigma_0^2}{2} \left[1 + \left( \frac{z}{k\sigma_0^2} \cdot \frac{e^{2\Lambda} - 1}{2\Lambda} \right)^2 \right].\tag{16c}
\setcounter{equation}{16}
\label{eq16}
\end{align}
The results demonstrate that the expectation value of particle position $\langle x \rangle$ remains invariant, while the position uncertainty $\Delta x$ is modified by the acceleration effects. Next, we perform the inverse Fourier transform to derive the momentum-space wave function as
\begin{align}
\Psi(p,z) &= \frac{1}{\sqrt{2\pi\hbar}} \int \Psi(x,z) e^{-i px/\hbar} \mathrm{d}x \nonumber \\
          &= \frac{\exp\left\{ -\frac{p^2 \sigma_0^2}{2\hbar^2} \left(1 + \frac{iz}{k\sigma_0^2} \cdot \frac{e^{2\Lambda} - 1}{2\Lambda}\right) - \frac{i p x_0}{\hbar} \right\}}{(\pi/ \sigma_0^2)^{1/4} \sqrt{\hbar}} .
          \label{eq17}
\end{align}
Differing from Eq.~\eqref{eq14} in both amplitude and phase, the effects of acceleration on the momentum-space wave function of Eq.~\eqref{eq17} persist only in phase correlations. The reason why it affects position-space and not the momentum-space representation wavefunction is somehow linked to the geometry of Rindler spacetime, supplemented by field-theoretic constraints for Gaussian states. Rindler space has no Ricci curvature (due to vanishing Riemann tensor), and  the corresponding  transformation involves position-dependent scaling of time and space, resulting in acceleration-modified position wave function by affecting Gaussian width (via real terms) and phase (via imaginary terms). The momentum-space counterpart, being Fourier transformed position-space wavefunctions, involves modulated phase factors only, while retaining its Gaussian amplitude. This may be a consequence of unitary time evolution dynamics of the field preserving probability densities. As we show, this also eliminates any contribution from acceleration for momentum uncertainty. The expectation value of the particle’s momentum and its momentum uncertainty can be written as 
\begin{align}
\langle p \rangle &= 0 ,
\tag{18a} \\
\langle p^2 \rangle &= \frac{\hbar^2}{2\sigma_0^2} ,
\tag{18b} \\
(\Delta p)^2 &= \frac{\hbar^2}{2\sigma_0^2}.
\tag{18c}
\setcounter{equation}{18}
\label{eq18}
\end{align}
The expressions above thus entail no impact of acceleration on the momentum, and therefore correspond exactly to that of flat spacetime observations. Furthermore, the position-momentum uncertainty relation in this framework turns out to be 
\begin{equation}
(\Delta x)^2 (\Delta p)^2 = \frac{\hbar^2}{4} \left[ 1 + \left( \frac{z}{k\sigma_0^2} \cdot \frac{e^{2\Lambda} - 1}{2\Lambda} \right)^2 \right].
\label{eq19}
\end{equation}
Hence, based on the previously conjectured duality between time evolution in Feynman approach and spatial diffraction of the classical field in Collins formalism, the Rindler observer only perceives modifications to the temporal dependence of uncertainty relation. Consequently, this manifests as an accelerated dispersion of Gaussian wavepackets. Importantly, this does not imply that the momentum uncertainty \(\Delta p\) remains entirely unchanged. As shown in Eq.~\eqref{eq19}, any variation in the position uncertainty \(\Delta x\) necessarily entails a corresponding change in \(\Delta p\). What we rather mean here is that an explicit acceleration-dependence of $\Delta p$ is not manifested. However, the original spirit of uncertainty relation is still in-tact. It would not be unfair to suggest here that the acceleration effects induce some sort of nonlocality in the spatial domain through Eq.~\eqref{eq19} or in the temporal domain, owing to the aforementioned space and time correspondence. Furthermore, we note that the limit ($a\rightarrow0$) giving
\begin{equation}
(\Delta x)^2 (\Delta p)^2 \approx  \frac{\hbar^2}{4} \left[ 1 + \left( \frac{z}{k\sigma_0^2}  \right)^2 \right],
\label{eq20}
\end{equation}
which fully conforms to the inertial flat spacetime results \cite{blinder1968}. Fig.~\ref{Fig1} illustrates the evolution of the position-momentum uncertainty relation as a function of propagating distance. The red dashed curve represents the uncertainty in flat space, while the solid curves correspond to accelerated frames with varying $\Lambda$ values. Note that $\Lambda$ is dimensionless, and critically governs the dispersion dynamics. Positive $\Lambda$ enhances the wave packet delocalization, accelerating the uncertainty growth rate, while negative $\Lambda$ suppresses quantum spread, decelerating the dispersion rate. This dichotomy arises from the acceleration-induced spacetime curvature in Rindler frame. Such properties are also similar to the classical beams propagating in accelerated space, such as Airy beams and hollow beams \cite{ding2021b,wang2023}.   
\section{Numerical simulations and cosmological aspects}
\subsection{Energy density distribution and cosmological redshift}
It is well-established that accelerated frames induce Unruh effect, a phenomenon sharing the same physical origin as Hawking radiation from black holes \cite{hawking1975}: thermality of the quantum vacuum arising out of disagreement of the particle content of the field involving various non-inertial observers. Building upon the free-particle wave function derived in Rindler spacetime (Sec.~\ref{sec2}), we now draw an analogy with $\Lambda$CDM model to explore potential cosmological implications.\\
\indent Assuming that the quantized optical field in the accelerated frame behaves like a photon gas obeying a Bose-Einstein distribution, the mean energy density of the free-particle can be expressed as
\begin{equation}
U = \frac{\hbar \omega'}{e^{\beta \hbar \omega'} - 1}.
\label{eq21}
\end{equation}
Here, $\beta = 1/ (k_B T)$, $k_B$ is the Boltzmann constant and $T$ denotes temperature. In one-dimensional space, the mode density $(\omega', \omega' + \mathrm{d}\omega')$ is given by $d_k = 4/\lambda^2$. Building on the relation derived earlier in Eq.~\eqref{eq9}, i.e., $\omega' = v'k/2 = v'\pi/\lambda$ ($v'$ represents the group velocity of light in the accelerated frame), the Planck energy density distribution as a function of wavelength $\lambda$ can be expressed as
\begin{align}
\rho(\lambda) &= d_k U \nonumber \\
              &= \frac{4\hbar\pi v'}{\lambda^3 \left(e^{\beta\hbar \pi v'/\lambda} - 1\right)} \nonumber \\
              &= \frac{4\hbar\pi v \cdot \frac{e^{2\Lambda} - 1}{2\Lambda}}{\lambda^3 \left( \exp\left\{ \frac{\beta\hbar \pi v}{\lambda} \cdot \frac{e^{2\Lambda} - 1}{2\Lambda} \right\} - 1 \right)}.
              \label{eq22}
\end{align}
The Eq.~\eqref{eq22} clearly indicates that the energy density distribution of the electromagnetic field bears acceleration effects through the parameter $\Lambda$, and can be plotted to reveal the modified thermal energy density distribution of the electromagnetic field. 
   \begin{figure}[htbp]
   \centering
   \includegraphics[width=1\linewidth]{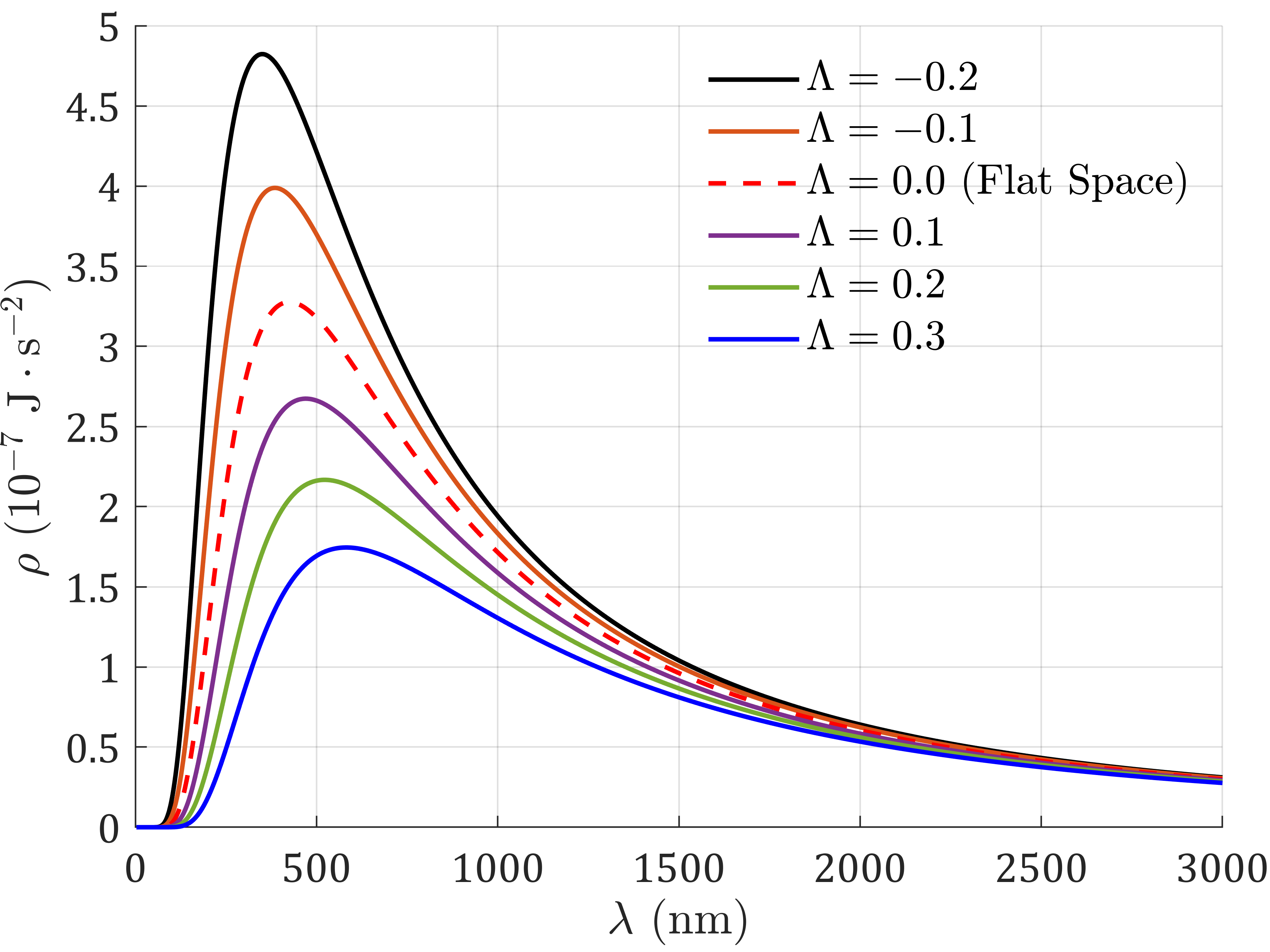}
      \caption{Spectral energy density distributions as a function of wavelength. Parameters included $T =6000~\mathrm{K}$, $v = c$.}
         \label{Fig2}
   \end{figure}

As shown in Fig.~\ref{Fig2}, two distinct regimes appear. One with positive $\Lambda$, which is characterized by the observer moving in the direction of the field propagation, while negative $\Lambda$ indicates the opposite direction. At the outset, the plots reveal an enhancement in peak energy density with positive $\Lambda$, while at the same time the peak of the curves shifts towards high frequency (low $\lambda$) end. This behavior undoubtedly parallels that of spectral energy distribution in the conventional Planck distribution. However, the role of temperature in conventional picture is taken up by the acceleration in our model. This is expected since the thermal effects in the Rindler frame arise out of acceleration, as is typically associated with Unruh effect \cite{unruh1976}, with a temperature estimate of 
\begin{equation}
T_{\mathrm{U}} = \frac{\hbar a}{2\pi k_{\mathrm{B}} c},
\label{eq23}
\end{equation}
and this fact puts acceleration and temperature on same footing. Likewise, negative $\Lambda$ exposes the observer to a less energy density associated with the field. Such situation with a suppressed flux would be tantamount to so-called anti-Unruh effect \cite{brenna2016}, wherein detectors coupled to field, under certain specific conditions, would cool down in contrast to heating mechanism expected from conventional Unruh effect. However, we caution that the picture presented here for cooling effect is contingent on the sense of direction of accelerated observer, whereas the case is \textit{not} so with standard anti-Unruh effect reported in the literature.  

For computing the peak energy density $\rho$ corresponding to a certain critical wavelength, say $\lambda_{\text{max}}$, we set $(\frac{\mathrm{d}\rho (\lambda)}{\mathrm{d}\lambda})_{\lambda=\lambda_{\text{max}} }=0$, which after the substitution $\beta\hbar \pi v'/\lambda = y'$ yields the expression as
\begin{equation}
-\frac{12\beta y'}{\lambda^3 (e^{y'} - 1)} + \lambda y'\frac{ 4\beta y' e^{y'}}{\lambda^4 (e^{y'} - 1)^2} = 0.
\label{eq24}
\end{equation}
The solution of Eq.~\eqref{eq24} is related to Lambert function $W (\cdot)$ and can be expressed as
\begin{equation}
\lambda_{\text{max}} = \frac{\hbar \pi v}{k_{\mathrm{B}} T \bigl[3 + W(0,-3e^{-3})\bigr]} \cdot \frac{e^{2\Lambda} - 1}{2\Lambda}.
\label{eq25}
\end{equation}
That's consistent with the Wien's displacement law \cite{wien1896}. It also helps us to give an estimate of acceleration-induced redshift as 
\begin{equation}
\alpha = \frac{\lambda_{\text{max}} - \lambda_{\text{max}0}}{\lambda_{\text{max}0}} 
= \frac{e^{2\Lambda} - 1}{2\Lambda} - 1 > -1.
\label{eq26}
\end{equation}
Here the $\lambda_{\text{max}}$ represents the critical wavelength from the Rindler observer’s  perspective, while the $\lambda_{\text{max}0}$ is from the flat space observer’s perspective. As seen from the above expression, the redshift arises directly out of accelerated motion of the observer along with the bound $\alpha > -1$ that guarantees a positive redshift, which is in agreement with the very rampant positive redshift found in the Universe. 
   \begin{figure}[htbp]
   \centering
   \includegraphics[width=1\linewidth]{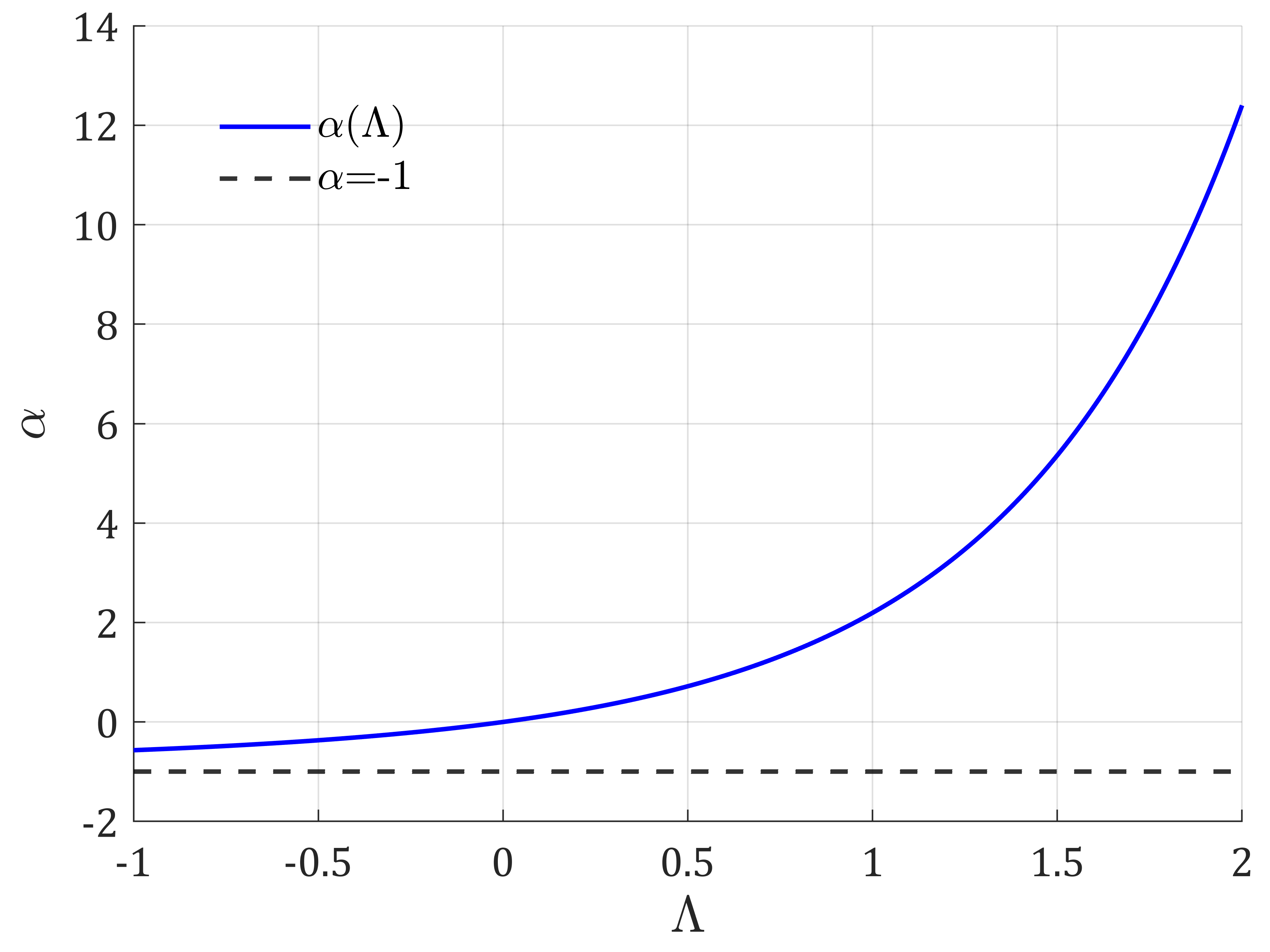}
      \caption{The dependence of the redshift $\alpha$ on the parameter $\Lambda$. The dashed line indicates the boundary condition, when $\alpha = -1$ .}
         \label{Fig3}
   \end{figure}

From a mathematical point of view, Eq.~\eqref{eq26} plotted in Fig.~\ref{Fig3} reveals that the redshift $\alpha$ and parameter $\Lambda$ exhibit a single-valued, exponential relationship, with $\alpha$ always satisfying $\alpha > -1$ and passing through the origin ($\Lambda$ = 0, $\alpha$ = 0). Furthermore, the inverse function of Eq.~\eqref{eq26} allows solving for $\Lambda$ as a function of $\alpha$,
\begin{align}
\Lambda(\alpha) &= -\frac{ \frac{1}{1+\alpha} + W\left(-1, -\frac{e^{-1/(\alpha+1)}}{\alpha+1}\right) }{2}.
\label{eq27}
\end{align}
The analytical solution for Eq.~\eqref{eq27} is valid only for $\alpha> 0$ (positive redshift), reflecting the observational dominance of cosmic expansion. Ultimately, we get a unified equation linking redshift $\alpha$, propagation distance $z$, and acceleration $a$ in the form:
\begin{equation}
\frac{a z}{c^2} = \Lambda(\alpha) = -\frac{ \frac{1}{1+\alpha} + W\left(-1, -\frac{e^{-1/(\alpha+1)}}{\alpha+1}\right) }{2} .
\label{eq28}
\end{equation}
The result is in some sense an interesting finding, with the scope to interpret the cosmological redshift as observed in terrestrial settings being a consequence of accelerated motion of observer due to some mysterious force, called dark energy. Dark energy hypothesis is one of the very profound explanations to understand the observed accelerated expansion of the Universe \cite{peebles2003,riess1998,Perlmutter1999}. Now, recall that the well-known cosmological redshift of galaxies or the faint glow of Cosmic Microwave Background radiation is a direct manifestation of stretching of spacetime in Big Bang cosmology and that the presence of radiation emanates from the decoupling of initial radiation from the matter components of the Universe \cite{peebles2020}. In this regard, above results suggest the possibility of an additional acceleration-driven redshift to the already known Big Bang mechanism, which obviously is the case with dark energy model. Somewhat surprising is that, in our model, there is no need of radiation that will experience redshift as is the case with modern cosmology; the apparent redshift might be rather some thermal radiation sourced by pure vacuum state of field, thanks to the the workings of the Unruh effect. This somehow mimics the de Sitter radiation in a spacetime purely characterized by dark energy (positive cosmological constant), pictured by Gibbons-Hawking effect \cite{gibbons1977}.

\subsection{The equivalent acceleration}
In $\Lambda$CDM model, the geometry of a homogeneous and isotropic expanding Universe is described by the FLRW metric, which in spherical coordinates takes the form \cite{young2017}:
\begin{equation}
\mathrm{d}s^2 = c^2 \mathrm{d}t^2 - a_s^2 [ \frac{\mathrm{d}r^2}{1 - k_s r^2} + r^2 ( \mathrm{d}\theta^2 + \sin^2\theta \, \mathrm{d}\phi^2 ) ].
\label{eq29}
\end{equation}
In FLRW metric, $a_s(t)$ is the scale factor of the Universe, describing the expansion or contraction of the Universe over time. Here, $k_s$ is the curvature constant, when $k_s=1$, it corresponds to a closed Universe, while $k_s$ = 0 to a flat Universe, and $k_s=-1$ to an open Universe. Meanwhile, $t$ is the time coordinate, and $(r, \theta, \phi)$ are the spatial coordinates. Starting with the FLRW metric, the comoving distance can be deduced in $\Lambda$CDM model, and it represents the distance measured between an observer and a celestial object as they both expand with the Universe \cite{peebles2003}. When neglecting radiation and curvature, the definition of the comoving distance can be expressed by \cite{young2017}
\begin{equation}
\mathcal{D}_M = \frac{c}{H_0} \int_{0}^{\alpha} \frac{\mathrm{d}\alpha'}{\sqrt{\Omega_m (1 + \alpha')^3 + \Omega_{\Lambda}}},
\label{eq30}
\end{equation}
where $(\Omega_m, \Omega_\Lambda)$ represent the matter and dark energy density parameter, $H_0$ represents the Hubble constant. In reality, cosmic expansion induces spectral redshift in observed light. The luminosity distance $\mathcal{D}_L$, which accounts for redshift effects caused by the Universe's expansion, is related to the comoving distance through \cite{young2017}
\begin{equation}
\mathcal{D}_L = (1 + \alpha) \mathcal{D}_M.
\label{eq31}
\end{equation}
It is interesting to note that, in principle, the propagation distance $z$ involved in the Rindler description of  light field can take arbitrary values. However, for the sake of quantitatively appreciating the physics behind these observations, we consider set of three distinct cases depending on some explicit  values of propagation distance.    

\subsubsection{Acceleration predicted from the luminosity distance}\label{case1}
Here, we postulate that the light propagation distance in Rindler spacetime is equivalent to the luminosity distance $\mathcal{D}_L$ as 
\begin{equation}
z=\mathcal{D}_L.
\label{eq32}
\end{equation}
By coupling with Eq.~\eqref{eq30}, we derive the equivalent acceleration $a$ in Eq.~\eqref{eq28} as
\begin{equation}
\frac{a}{c H_0} = -\frac{ \frac{1}{1+\alpha} + W\left(-1, -\frac{e^{-1/(\alpha+1)}}{\alpha+1}\right) }{ 2(1+\alpha) \int_0^\alpha \mathrm{d}\alpha'/\sqrt{\Omega_m (1+\alpha')^3 + \Omega_\Lambda}} .
\label{eq33}
\end{equation}
   \begin{figure}[htbp]
   \centering
   \includegraphics[width=1\linewidth]{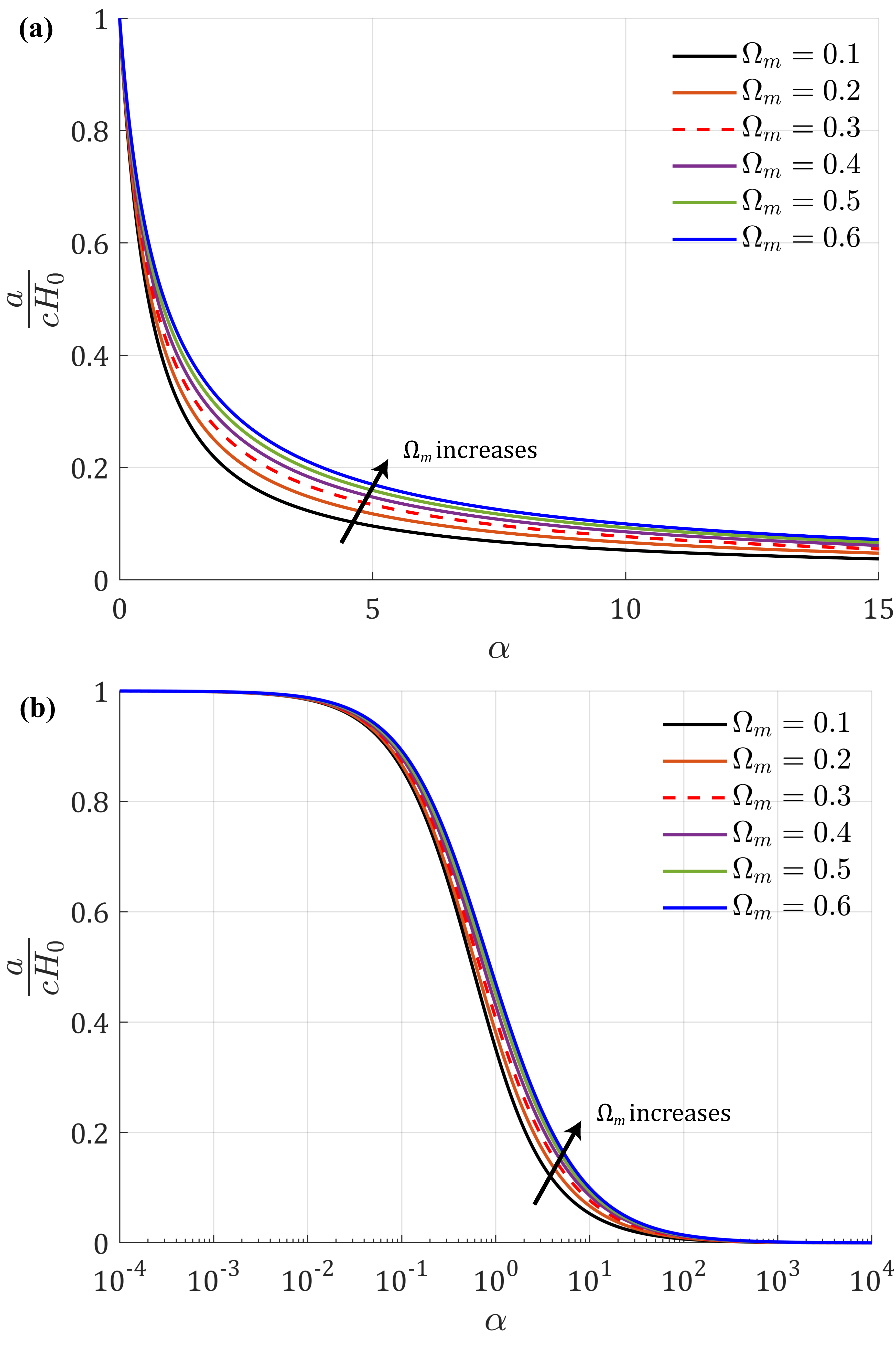}
      \caption{Relationship between redshift and equivalent acceleration in (a) the normal and (b) logarithmic scales of $\alpha$. The curves are calculated from Eq.~\eqref{eq33} under increasing \(\Omega_m\).}
         \label{Fig4}
   \end{figure}

In Fig.~\ref{Fig4}(a), the curves depict the equivalent acceleration $a$ of galaxies derived from our model with varying matter density parameters $\Omega_m$ ($\Omega_m + \Omega_\Lambda = 1$). The equivalent accelerations are normalized by $c H_0$. Different density parameters lead to distinct evolutionary paths for the equivalent acceleration. A small matter density parameter will cause the equivalent acceleration to be more sensitive to redshift. This plot demonstrates that cosmic equivalent acceleration is not constant. The acceleration was slower in the early Universe, and it increases as the Universe evolves. However, when the redshift approaches zero ($\alpha$ = 0), instead of tending to zero, the acceleration will converge to a constant value $c H_0$. To clearly appreciate the convergence of acceleration for the extremities of $\alpha$, we plot Fig.~\ref{Fig4}(b) as on a logarithmic scale and a wide range of redshifts. The figure illustrates that as $\alpha \rightarrow 0$ (the local Universe limit), the equivalent acceleration approaches a constant value $c H_0$ as earlier.  

\subsubsection{Acceleration predicted from the comoving distance}\label{case2} 
\begin{figure}[htbp]
   \centering
   \includegraphics[width=1\linewidth]{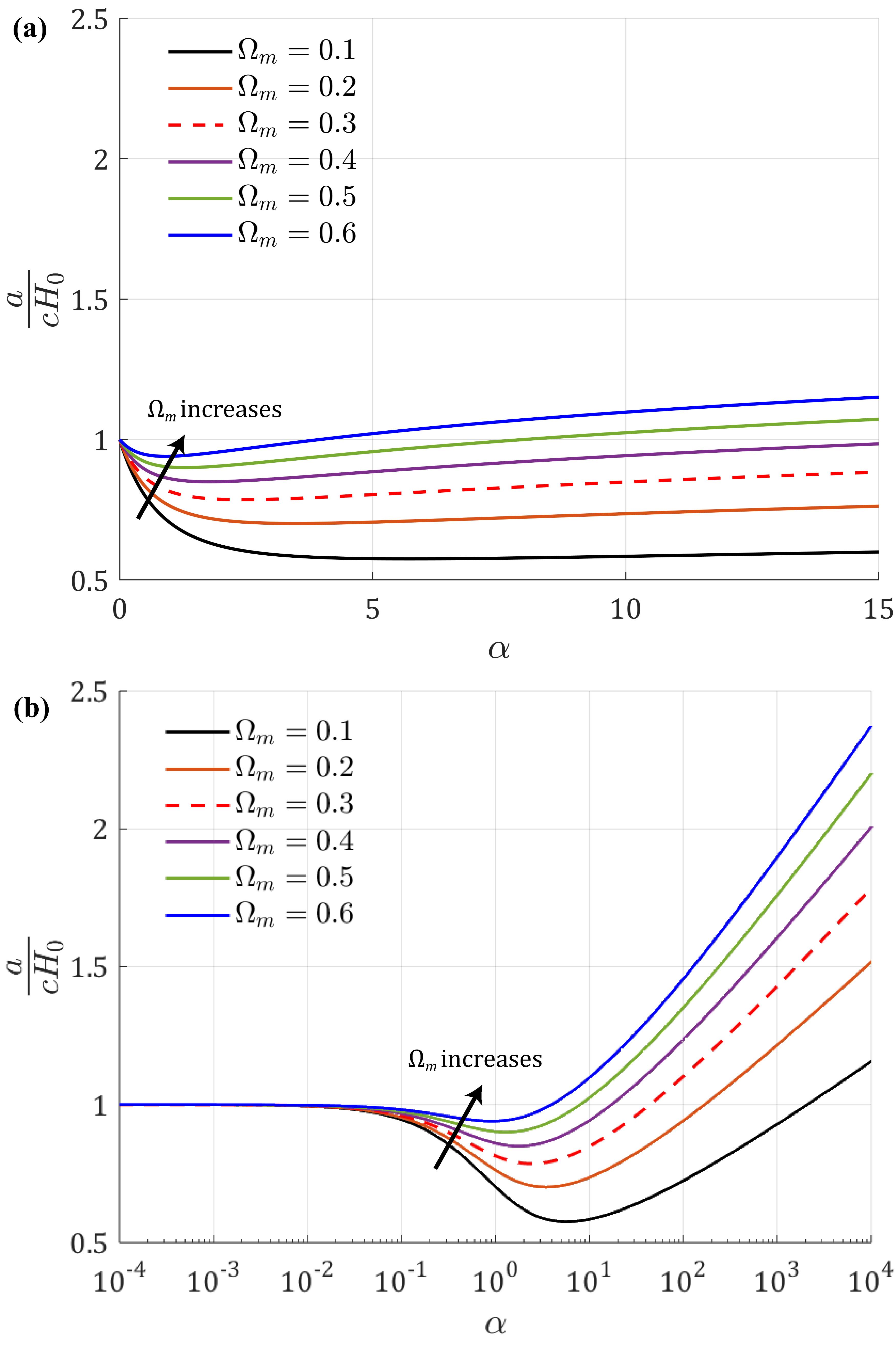}
      \caption{Relationship between redshift and equivalent acceleration in (a) the normal and (b) logarithmic scales of $\alpha$. The curves are calculated from Eq.~\eqref{eq35} under increasing \(\Omega_m\).}
         \label{Fig5}
   \end{figure}

One also has the liberty of defining equivalent acceleration  in terms of comoving distance $\mathcal{D}_M$.  Starting from the relation
\begin{equation}
z=\mathcal{D}_M.
\label{eq34}
\end{equation}
and coupling with Eq.~\eqref{eq28}, we derive the equivalent acceleration $a$  as
\begin{equation}
\frac{a}{c H_0} = -\frac{ \frac{1}{1+\alpha} + W\left(-1, -\frac{e^{-1/(\alpha+1)}}{\alpha+1}\right) }{ 2 \int_0^\alpha \mathrm{d}\alpha'/\sqrt{\Omega_m (1+\alpha')^3 + \Omega_\Lambda}},
\label{eq35}
\end{equation}
helping us to alternatively express the results in Fig.~\ref{Fig4} in terms of Fig.~\ref{Fig5}, respectively. The Fig.~\ref{Fig5} demonstrate some agreement with the results to that of luminosity distance: the equivalent acceleration also approaches $c H_0$ for local Universe limit ($\alpha \rightarrow 0$), while it diverges for early Universe limit ($\alpha \rightarrow \infty$). In contrast to case \hyperref[case1]{1}, we can see clearly there are turning points in Fig.~\ref{Fig5}(b). These turning points indicate that the Universe has undergone phases where the equivalent acceleration changes its course of action from decreasing to increasing values with respect to a Rindler observer. This might resonate with the inflationary paradigm where the rapid (exponential) expansion of the Universe is believed to have been caused by some form of energy source.

\subsubsection{Acceleration predicted from the relativistic redshift}\label{case3}  
In this case, we begin with the relativistic frequency shift,
\begin{equation}
\frac{\nu_R^2}{\nu_F^2} = \frac{c - u}{c + u},
\label{eq36}
\end{equation}
where $\nu_R$ represents the frequency of observer in Rindler spacetime, $\nu_F$ in flat spacetime, and $u$ corresponds to the instantaneous velocity of observer in Rindler spacetime at instantaneous time. By applying the relationship $\lambda\nu = c$, we deduce the redshift as
\begin{equation}
\alpha = \frac{\lambda_R}{\lambda_F} - 1 = \sqrt{\frac{c + u}{c - u}} - 1,
\label{eq37}
\end{equation}
where $\lambda_R$ and $\lambda_F$ represent the wavelengths for the observer in Rindler and flat spacetime, respectively. Combining this result with Hubble’s law \cite{hubbble1929}, we obtain
\begin{equation}
H_0 d = u,
\label{eq38}
\end{equation}
we obtain the distance as
\begin{equation}
d = \frac{\alpha (2 + \alpha) c}{H_0 (2 + 2\alpha + \alpha^2)}.
\label{eq39}
\end{equation}
Then we define the light propagation distance in Rindler spacetime in terms of Hubble’s Law, which reads
\begin{equation}
z = d.
\label{eq40}
\end{equation}
By coupling with Eq.~\eqref{eq28}, we derive the equivalent acceleration $a$  as 
\begin{equation}
\frac{a}{c H_0} = \frac{ \frac{1}{1+\alpha} + W\left(-1, -\frac{e^{-1/(\alpha+1)}}{\alpha+1}\right)  }{ \frac{-2\alpha (2 + \alpha)}{2 + 2\alpha + \alpha^2} }.
\label{eq41}
\end{equation}
   \begin{figure}[htbp]
   \centering
   \includegraphics[width=1\linewidth]{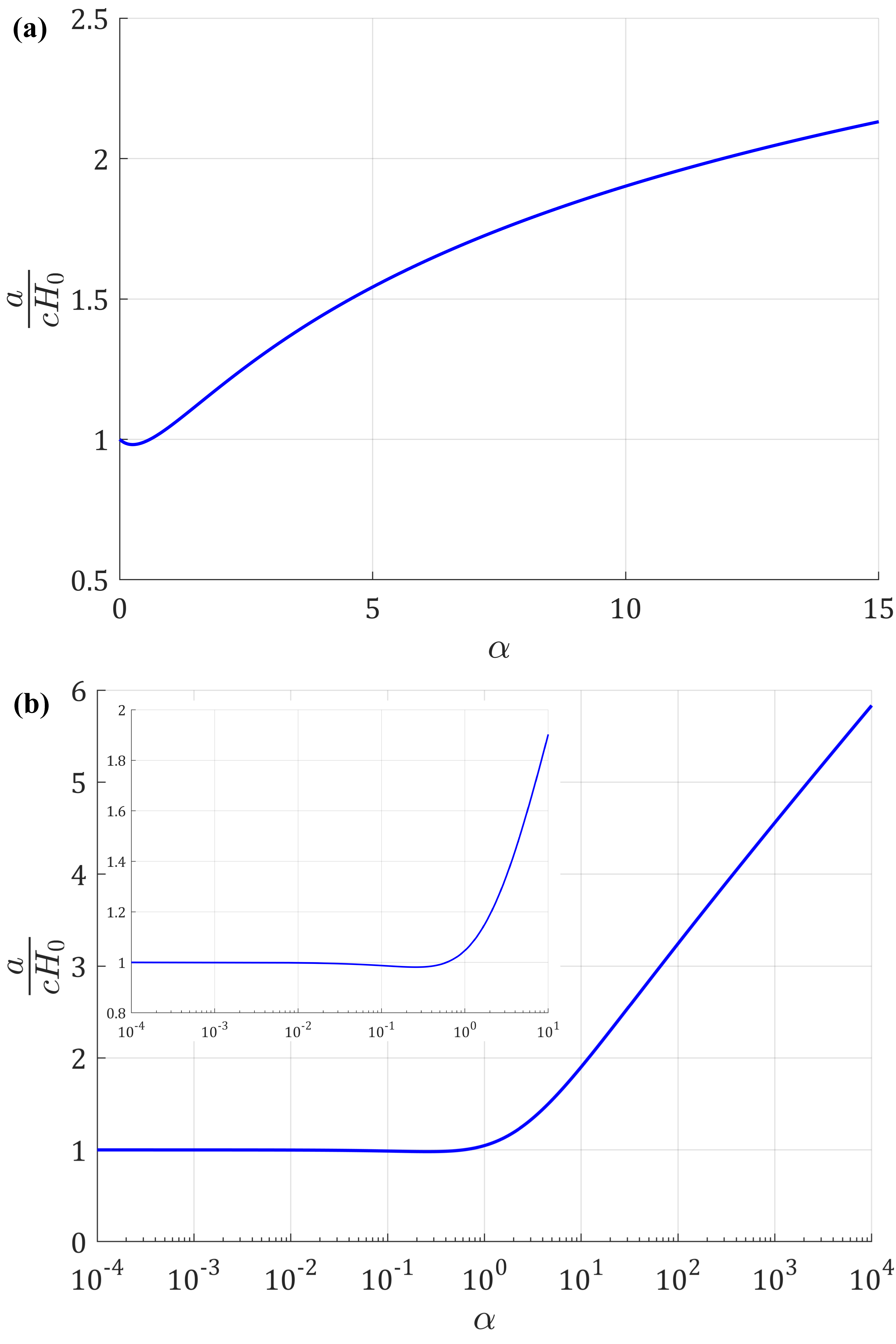}
      \caption{Relationship between redshift and equivalent acceleration in (a) the normal and (b) logarithmic scales of $\alpha$. The curves are calculated from Eq.~\eqref{eq41}.}
         \label{Fig6}
   \end{figure}

In terms of current and early epochs of the Universe, Fig.~\ref{Fig6} shows similar characteristics to that of Fig.~\ref{Fig5}. Explicitly speaking, the quantity $\frac{a}{c H_{0}}$ converges to same value for $\alpha=0$. The turning point in this case tends to move towards the origin compared to Fig.~\ref{Fig5}(b). We have demonstrated this more clearly in the inset picture of Fig.~\ref{Fig6}(b) which is on a logarithmic scale. The similarity in the behavior of Fig.~\ref{Fig5} and Fig.~\ref{Fig6} is somewhat telling because the values of the equivalent acceleration plotted in Fig.~\ref{Fig6} were calculated from relativistic frequency shift-- \textit{not} the $\Lambda$CDM model.  

\section{Conclusion} 
Now, as is well-established that the Universe is currently undergoing a phase of accelerated expansion \cite{riess1998,Perlmutter1999}, the exact magnitude of acceleration is topic of great attention as further refinements and adjustments of the values of various cosmological parameters are being actively performed. In this context, our model predicts accurate estimates of the acceleration, which has been noted earlier from extremal values of redshift parameter $\alpha\in(0, \infty)$ corresponding to current and early epochs of the Universe, respectively. It indicates that the cosmic acceleration initiates shortly after Big Bang and continues to grow and saturates at some current epoch, somehow resonating with dynamical dark energy model involving a (non-constant) time-dependent cosmological constant \cite{copeland2006,zhao2017}. These findings may have immediate implications for constraining cosmological parameters related to expansion dynamics. \\
\indent Certain important points are in order. The usual homogeneous and isotropic picture of FLRW cosmological expansion is somehow an approximation positing a uniform inertial frame-like dynamics of the expansion and the observations related to it. It involves real stretching of spacetime of the Universe. On the other hand, the key ingredient in our work is the Rindler description of the field evolution, in which redshift phenomena are solely attributable to non-inertial type motion of observers in line with Unruh effect without involving any stretching of spacetime. Despite this, both frameworks give remarkably similar description of various aspects of the field dynamics.  This correspondence may be insightful for simulations of cosmic dynamics in future analogue gravity program. \\
\indent Having discussed the three cases, the question that might naturally arise in one's mind is whether the picture presented here corresponds to the standard big bang cosmology where one would expect high values of acceleration in the early and late epochs of the Universe corresponding to, in particular, the inflationary cosmology and the late time acceleration, respectively. Indeed that is roughly the situation here for the case \hyperref[case2]{2} and case \hyperref[case3]{3}. For the case \hyperref[case1]{1}, since the equivalent acceleration decays asymptotically to zero as $\alpha\rightarrow \infty$, one can not recover the dynamics of the Universe, given our knowledge of the well-known observations in standard cosmology. The late time acceleration of the Universe  driven by dark energy is also  manifested here as the equivalent acceleration starts increasing for lower values of redshift. In particular, for the case \hyperref[case1]{1}, the rise in equivalent acceleration occurs monotonically, while as it passes through turning points for case \hyperref[case2]{2} and case \hyperref[case3]{3}. \\
\indent In summary, we proposed a model of light evolution from the perspective of accelerated observers in Rindler frame using two important methodologies, Collins diffraction formula and Feynman path integral approach, respectively providing classical and quantum descriptions of the light field. The approach can help us to assume field like a wavepacket describing a free particle wave function from which we compute acceleration-induced corrections to quantum uncertainty principle. We showed that, while the core traits of position-momentum uncertainty relation are preserved, spatial part of the uncertainty relation suffers modifications while the momentum part remains unaffected. This represents a kind of enhanced nonlocal effects in the traditional uncertainty relation. Later, we computed the impact of acceleration on the Planckian spectral energy density distribution of the field and found enhancement (suppression) in the peak energy densities corresponding to positive (negative) values of acceleration parameter. Along with shifting of peaks of the distribution curves, it indicated an equivalence between temperature and acceleration, similar to Unruh effect.\\   
\indent Furthermore, we investigated its relation to the well-known cosmological redshift, and attempted to assess the consequences of our setup by computing an equivalent acceleration in three different cases. The equivalent acceleration  approaches $c H_0$ for the local Universe limit in all cases. Concerning  second and third case, we see our results being in harmony with the well-known results in standard cosmology involving expansion and related parameters, while first case seems inapt, especially for early epochs of the Universe. Some intriguing results concerning the emergence of cosmic acceleration vis-\`{a}-vis redshift factor were obtained, which may of great significance to emulating cosmological phenomena in lab settings. 

\acknowledgments
      This work is supported by the National Natural Science Foundation of China (NSFC) (grants No.62375241).
\bibliography{wfj}
\end{document}